\documentclass[%
 aip,
 jcp,%
 amsmath,amssymb,
 reprint,%
]{revtex4-1}

\usepackage{graphicx}         
\usepackage{dcolumn}          
\usepackage{bm}               

\begin{document}

\title{Significant contributions of Albrecht's $\bm{A}$ term to
\textit{non}-resonant Raman scattering processes}

\author{Zu-Yong Gong}%
\affiliation{Hefei National Laboratory for Physical Science at the
Microscale, Department of Chemical Physics, School of Chemistry and
Materials Science, University of Science and Technology of China, Hefei,
230026 Anhui, P. R. China.}
\affiliation{Department of Theoretical Chemistry and Biology, School of
Biotechnology, Royal Institute of Technology, S-106 91 Stockholm,
Sweden.}%

\author{Guangjun Tian}%
\affiliation{Department of Theoretical Chemistry and Biology, School of
Biotechnology, Royal Institute of Technology, S-106 91 Stockholm,
Sweden.}

\author{Sai Duan}%
\email{said@theochem.kth.se}%
\affiliation{Department of Theoretical Chemistry and Biology, School of
Biotechnology, Royal Institute of Technology, S-106 91 Stockholm,
Sweden.}

\author{Yi Luo}%
\affiliation{Hefei National Laboratory for Physical Science at the
Microscale, Department of Chemical Physics, School of Chemistry and
Materials Science, University of Science and Technology of China, Hefei,
230026 Anhui, P. R. China.}%
\affiliation{Department of Theoretical Chemistry and Biology, School of
Biotechnology, Royal Institute of Technology, S-106 91 Stockholm,
Sweden.}%

\date{\today}

\begin{abstract}
The Raman intensity can be well described by the famous Albrecht
equation that consists of $A$ and $B$ terms. It is well known that the
contribution from Albrecht's $A$ term can be neglected without loss of
accuracy for far off-resonant Raman scattering processes. However, as
demonstrated in this study, we have found that this widely accepted
long-standing assumption fails drastically for totally symmetric
vibration modes of molecules in general off-resonant Raman scattering.
Perturbed first principles calculations for water molecule show that
strong constructive interference between the $A$ and $B$ terms occurs
for the Raman intensity of the symmetric O-H stretching mode, which can
account for about 40\% of the total intensity. Meanwhile, a minor
destructive interference is found for the angle bending mode. The state
to state mapping between the Albrecht's theory and the perturbation
theory allows us to verify the accuracy of the widely employed
perturbation method for the dynamic/resonant Raman intensities. The
model calculations rationalized from water molecule with the bending
mode show that the perturbation method is a good approximation only when
the absolute energy difference between the first excited state and the
incident light is more than five times of the vibrational energy in
ground state.
\end{abstract}

\maketitle

\section{Introduction}

Raman spectroscopy\cite{raman1928nature} is one of the standard
vibrational spectroscopic tools that has been extensively applied in
different fields.  The intensity of the Raman scattering can be
generally treated by Albrecht's Raman theory\cite{albrecht1961jcp}, in
which the polarizability is expanded into two terms, \textit{i.e.} the
famous Albrecht's $A$ and $B$ terms\cite{albrecht1961jcp,long2002}
\begin{equation}
\alpha_{pq}=A_{pq}+B_{pq},
\label{eq:alpha}
\end{equation}
where $p$ and $q$ represent Cartesian coordinates. If both initial and
final electronic states belong to the ground state, they can be written
as\cite{long2002,albrecht1961jcp}%
{\small%
\begin{equation}
\begin{split}
A_{pq}=&\frac{1}{\hbar}\sum_{r\neq{g}}\sum_{v^r=0}^\infty
\frac{p_{e^ge^r}^0q_{e^re^g}^0}{\omega_{e^rv^r:e^gv^i}-\omega}
\langle{v}^f|v^r\rangle\langle{v}^r|v^i\rangle\\
+&\frac{1}{\hbar}\sum_{r\neq{g}}\sum_{v^r=0}^\infty
\frac{p_{e^ge^r}^0q_{e^re^g}^0}{\omega_{e^rv^r:e^gv^f}+\omega}
\langle{v}^f|v^r\rangle\langle{v}^r|v^i\rangle\\
B_{pq}=&\frac{1}{\hbar}\sum_{r\neq{g}}\sum_{v^r=0}^\infty
\frac{p_{e^ge^r}^{k,0}q_{e^re^g}^0}{\omega_{e^rv^r:e^gv^i}-\omega}
\langle{v}^f|Q_k|v^r\rangle\langle{v}^r|v^i\rangle\\
+&\frac{1}{\hbar}\sum_{r\neq{g}}\sum_{v^r=0}^\infty
\frac{p_{e^ge^r}^{0}q_{e^re^g}^{k,0}}{\omega_{e^rv^r:e^gv^f}+\omega}
\langle{v}^f|Q_k|v^r\rangle\langle{v}^r|v^i\rangle\\
+&\frac{1}{\hbar}\sum_{r\neq{g}}\sum_{v^r=0}^\infty
\frac{p_{e^ge^r}^{0}q_{e^re^g}^{k,0}}{\omega_{e^rv^r:e^gv^i}-\omega}
\langle{v}^f|v^r\rangle\langle{v}^r|Q_k|v^i\rangle\\
+&\frac{1}{\hbar}\sum_{r\neq{g}}\sum_{v^r=0}^\infty
\frac{p_{e^ge^r}^{k,0}q_{e^re^g}^0}{\omega_{e^rv^f:e^gv^f}+\omega}
\langle{v}^f|v^r\rangle\langle{v}^r|Q_k|v^i\rangle,
\end{split}
\label{eq:ab}
\end{equation}}%
where $\hbar$ is the reduced Planck's constant, $|e^g\rangle$ and
$|e^r\rangle$ are electronic ground and excited states, $|v^i\rangle$
and $|v^f\rangle$ are the initial and final vibrational states of
$|e^g\rangle$ associated with frequency $\omega^g$, $|v^r\rangle$ is the
vibrational state of $|e^r\rangle$ associated with frequency $\omega^r$,
$\omega$ is the frequency of the incident light,
$\omega_{e^rv^r:e^gv^i}$ is the frequency difference between
$|e^r\rangle|v^r\rangle$ and $|e^g\rangle|v^i\rangle$, $p_{e^ge^r}^0$ is
the transition dipole moment between $|e^g\rangle$ and $|e^r\rangle$ at
equilibrium geometry ($Q_0$), $p_{e^ge^r}^{k,0}$ is the derivative of
$p_{e^ge^r}^0$ with respect to specific normal mode $Q_k$.  Here we
should emphasize that \textit{all} modes could contribute as
intermediate states in the Raman processes. In other words,
$|v^r\rangle$ represents all possible combination of all modes,
\textit{i.e.}, $|v^r\rangle=|v_1^rv_2^r\cdots{v}_N^r\rangle$, where $N$
is the number of vibrational modes. In some cases, the denominator in
Eq.~\ref{eq:ab} would be written in terms of vertical excitation energy
$\Delta{E}_{rg}$, which obeys the relationship
$\Delta{E}_{rg}=\hbar\omega_{e^r:e^g}+\lambda$. Here $\lambda$ is the
reorganization energy.\cite{marcus1956jcp}

The Albrecht's theory is commonly used to study the resonant Raman
scattering,\cite{kelley2008jpca} although it is a general theory for all
Raman processes\cite{long2002}. It is well known that, for strongly
dipole-allowed transitions, $A$ term is dominant in the resonant Raman
spectra.\cite{kelley2008jpca,silverstein2012jcp} It is noted that $A$
term does not contribute to non totally symmetric
modes.\cite{mchale1998} On the other hand, for weakly dipole-allowed
transitions, both the $A$ and $B$ terms could significantly contribute
to the Raman scattering of totally symmetric modes. In these cases, the
relative magnitude of them could only be determined by quantum chemical
calculations.\cite{santoro2011jctc,Silverstein2012jcp_1,liang2012jctc,egidi2014jctc}
It is also noted that the effects of higher-order terms were also
discussed in recent studies.\cite{egidi2014jctc}

\begin{figure}[!tb]
\begin{center}
\includegraphics[width=8.6cm]{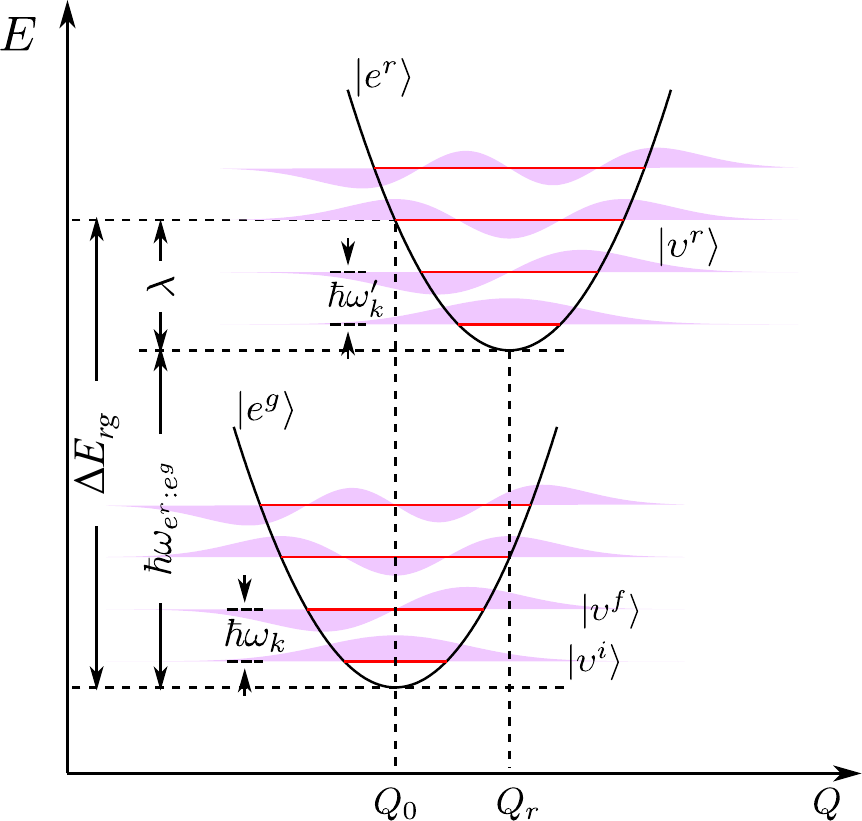}
\caption{\label{fig:model}Schematic drawing of the single displaced
harmonic model used to study the contribution of the $A$ term to the
Raman intensity. $Q_0$ ($\omega_k$) and $Q_r$ ($\omega_k^\prime$) are
the equilibrium geometries (vibrational frequencies) of ground state
$|e^g\rangle$ and excited state $|e^r\rangle$, respectively. All other
symbols can be found in Eq.~\ref{eq:ar}.}
\end{center}
\end{figure}

For \textit{non}-resonant Raman scattering processes, we focus on the
zero-frequency limit, \textit{i.e.}, the $\omega$ is neglected. In the
traditional treatment of the Albrecht's theory for \textit{non}-resonant
situations, it has been often assumed that
$\omega_{e^rv^r:e^gv^i}\approx\omega_{e^r:e^g}$.\cite{albrecht1961jcp}
As a result, the denominator in Eq.~\ref{eq:ab} is independent of the
vibrational state $|v^r\rangle$. Therefore, the summation of $v^r$ could
be calculated prior. Because of the completeness of $|v^r\rangle$ and
orthonormality between $|v^i\rangle$ and $|v^f\rangle$, the prior
summation will result in the $A$ term only responsible for the Raleigh
scattering and vanishing for \textit{non}-resonant Raman
scattering.\cite{albrecht1961jcp} Nowadays, this argument has been
widely accepted in the theory of \textit{non}-resonant Raman
scattering.\cite{mchale1998,long2002,smith2005,lombardi2008jpcc}
However, the assumption is apparently too strong. If we only consider
the fact that
$\Delta{E}_{rg}\gg\hbar\omega_{v^r}-\hbar\omega_{v^i}-\lambda$, in the
framework of harmonic approximation and expanding the denominator by
Taylor series, the component related to an \textit{arbitrary} excited
state $|e^r\rangle$ in Eq.~\ref{eq:ab} can be rewritten as
\begin{equation}
\begin{split}
A_{pq}^r=&\frac{2p_{e^ge^r}^0q_{e^re^g}^0}{\Delta{E}_{rg}}
\langle{v}^f|v^i\rangle\\
+&\frac{p_{e^ge^r}^0q_{e^re^g}^0}{\Delta{E}_{rg}^2}\left[\left(1+v^i+v^f\right)
\hbar\omega^g+2\lambda-\hbar\omega^r\right]\langle{v}^f|v^i\rangle\\
-&\frac{2p_{e^ge^r}^0q_{e^re^g}^0}{\Delta{E}_{rg}^2}
\sum_{v^r=0}^\infty\hbar
v^r\omega^r\langle{v}^f|v^r\rangle\langle{v}^r|v^i\rangle\\
B_{pq}^r=&\frac{2(p_{e^ge^r}^{k,0}q_{e^re^g}^0+p_{e^ge^r}^0q_{e^re^g}^{k,0})}
{\Delta{E}_{rg}}\langle{v}^f|Q_k|v^i\rangle\\
+&\mathcal{O}\left(\frac{\hbar\omega_{v^r}-\hbar\omega_{v^i}-\lambda}{\Delta{E}_{rg}}\right).
\end{split}
\label{eq:ar}
\end{equation}
Here we should emphasize the multimode nature of $\omega^g$ and
$\omega^r$. For instance, we have the expression of
$\hbar{v}^r\omega^r=\sum_{k=1}^N\hbar{v}_k^r\omega_k^r$. The schematic
drawings for a single vibrational mode of all definitions could be found
in Fig.~\ref{fig:model}, where $\omega_k$ and $\omega_k^\prime$ are the
vibrational frequencies related to $|e^g\rangle$ and $|e^r\rangle$,
respectively. The summation $A_{pq}^r$ over all $|e^r\rangle$ would
return to $A_{pq}$. If we consider the Stokes shift for the fundamental
frequency, \textit{i.e.} $v^i=0$ and $v^f=1$, the first two terms in
$A_{pq}^r$ vanish due to the orthogonality of the vibrational
wavefunctions in the ground state.  However, the last term survives and
should contribute to the \textit{non}-resonant Raman intensity.

The finite difference method\cite{hush1979cp} and coupled-perturbed
method\cite{amos1986cpl} have also been used to compute the intensity of
\textit{non}-resonant Raman spectra by directly differentiating the
electronic polarizabilities, \textit{i.e.}\cite{long2002,jcc-raman}
\begin{equation}
\alpha_{pq}=\langle{v}^f|\alpha_{e,pq}|v^i\rangle,
\label{eq:alpha1}
\end{equation}
where\cite{merzbacher1970}
\begin{equation}
\alpha_{e,pq}=\sum_{r\neq{g}}\frac{p_{e^ge^r}q_{e^re^g}}
{\Delta{E}_{rg}\pm\hbar\omega}.
\label{eq:ae}
\end{equation}
As a result, the component related to state $|e^r\rangle$ in
Eq.~\ref{eq:alpha1} can be rewritten as
\begin{equation}
\begin{split}
\alpha_{pq}^r
&=\frac{2p_{e^ge^r}^0q_{e^re^g}^0}{\Delta{E}_{rg}}\langle{v}^f|v^i\rangle\\
&-\frac{2p_{e^ge^r}^0q_{e^re^g}^0}{\Delta{E}_{rg}^2}
\frac{\partial\Delta{E}_{rg}}{\partial{Q}_k}\langle{v}^f|Q_k|v^i\rangle\\
&+\frac{2(p_{e^ge^r}^{k,0}q_{e^re^g}^0+p_{e^ge^r}^0q_{e^re^g}^{k,0})}
{\Delta{E}_{rg}}\langle{v}^f|Q_k|v^i\rangle.
\end{split}
\label{eq:ar1}
\end{equation}
Again, the zero-frequency limit is also applied in Eq.~\ref{eq:ar1} and
the summation $\alpha_{pq}^r$ would become $\alpha_{pq}$. In both cases,
the Raman cross section can be calculated from the polarizabilities in
Eqs.~\ref{eq:alpha} and \ref{eq:alpha1}.\cite{jcc-raman}

In this work, we comprehensively investigate the relationship between
Eqs.~\ref{eq:ar1} and \ref{eq:ar}. Comparing Eqs.~\ref{eq:ar1} and
\ref{eq:ar}, we can immediately find that the $B_{pq}^r$ term
corresponds to the last term in $\alpha_{pq}^r$.  Thus, the focus of
current study is the relationship for $A$ term. We first address this
issue in the model system and show that the $A$ term indeed contributes
to final results for \textit{non}-resonant conditions. Then, the $A$
term in time frame is briefly discussed. Finally, we take water monomer
as a realistic example to show the importance of including the $A$ term
for final results.

\section{Model System}

The first hint for the corresponding relationship for $A$ term between
Eqs.~\ref{eq:ar1} and \ref{eq:ar} can be found in the case of
$\partial{\Delta{E}_{rg}}/\partial{Q}_k=0$. If we consider that both
potential energy surfaces (PESs) of $|e^g\rangle$ and $|e^r\rangle$ are
harmonic, this case is equivalent to the situation that the displacement
between the two PESs is 0 ($\Delta{Q}=0$). We can immediately obtain
that the second term in $\alpha_{pq}^r$ is 0. On the other hand, because
of the no shift in PESs, either $\langle1|v^r\rangle$ or
$\langle{v}^r|0\rangle$ will be 0 due to the parity
symmetry.\cite{mchale1998} As a result, the last $A_{pq}^r$ term is also
0 for Raman scattering. Hence, one could notice that the last term in
$A_{pq}^r$ may correspond to the second term in $\alpha_{pq}^r$. It is
well known that the case can be realized in the non totally symmetric
vibrations.\cite{mchale1998} For such kind of vibrations, there always
exists at least one symmetric operator that makes the geometries of
$Q_0\pm\delta{Q}_k$ to be identity and then
$\partial{\Delta{E}_{rg}}/\partial{Q}_k=0$. Thus, the reason of $A$ term
has no contribution to the \textit{non}-resonant Raman intensity of non
totally symmetric vibrations is the orthogonality of vibrational
wavefunctions between ground and excited states rather than the
orthogonality of two vibrational wavefunctions in the ground state. The
former reason is exactly the same as that for the selection rule in
resonant Raman processes.\cite{mchale1998,smith2005,kelley2008jpca} It
is worth to note that this conclusion can also be generalized to
resonant Raman intensities as well as general PESs (such as double
well\cite{lee1985jce}) of excited states.

To further confirm the corresponding relationship between the last term
in $A_{pq}^r$ and the second term in $\alpha_{pq}^r$, the condition of
$\partial{\Delta{E}_{rg}}/\partial{Q}_k\neq0$ should be enforced. In the
framework of linear coupling model (LCM)\cite{macak2000cpl} where it
assumes $\omega_k=\omega_k^\prime$, both terms can be calculated
analytically and give the same result, \textit{i.e.},
\begin{equation}
\frac{\sqrt{2\hbar\omega_k^3}p_{e^ge^r}^0q_{e^re^g}^0\Delta{Q}}{\Delta{E}_{rg}^2}.
\label{eq:A-lcm}
\end{equation}
When $\Delta{Q}>0$, this result is equal to
$2p_{e^ge^r}^0q_{e^re^g}^0\hbar\omega_k\sqrt{S}/\Delta{E}_{rg}^2$, where
$S$ is the Huang-Rhys factor\cite{huang-rhys}. We should emphasis that
the current treatment is consistent with Ting's work\cite{ting1968sa}
which could be traced back to Shorygin's treatment\cite{shorygin1947zfk}
in 1947. However, Ting's work was restricted to the LCM scheme and the
vibronic coupling (the $B$ term) has not been included.\cite{lee1979jcp}
Moreover, Ting's algorithm was only widely applied in the two-state
model\cite{castiglioni1996prb, castiglioni1997prb,chou2009tca} and
intrinsically pre-resonant
conditions\cite{yeung1975sa,castiglioni1996prb,castiglioni1997prb}. In
off-resonance conditions, Ting's A term was usually considered to be
very small and negligible.\cite{lee1979jcp,bishop1997prb}

When going beyond the LCM, with the help of the general sum rules for
Franck-Condon integrals,\cite{sadlej1970sa} we could obtain that the
last term in $A_{pq}^r$ and the second term in $\alpha_{pq}^r$ both
equal to
\begin{equation}
\frac{\sqrt{2\hbar}\omega_k^{\prime2}p_{e^ge^r}^0q_{e^re^g}^0\Delta{Q}}
{\sqrt{\omega_k}\Delta{E}_{rg}^2}.
\label{eq:A-gen}
\end{equation}
We have further performed numerical calculations for more general
condition of $\partial{\Delta{E}_{rg}}/\partial{Q}_k\neq0$.  The
considered displaced harmonic model is shown in Fig.~\ref{fig:model}.
For example, in the case of $\omega_k=1600~\textrm{cm}^{-1}$,
$\omega_{k}^{\prime}=1800~\textrm{cm}^{-1}$, and
$\Delta{Q}_{k}=Q_r-Q_0=10~\textrm{a.u.}$, both terms give the same value
of $0.0055704918$ in the unit of
$2p_{e^ge^r}^0q_{e^re^g}^0\hbar/\Delta{E}_{rg}^2$. The analytical and
perfect numerical agreement not only confirm the equivalence of the two
terms but also show that, for modes with
$\partial{\Delta{E}_{rg}}/\partial{Q}_k\neq0$, the $A$ term can have
non-zero contribution to the \textit{non}-resonant Raman intensity. Note
that, here the corresponding $S$ is around 0.4,\cite{huang-rhys-2012}
which is reasonable for realistic molecules. Because the $|e^r\rangle$
is arbitrary in above discussions, a state to state relationship between
Albrecht's theory and perturbation method is thus established for
\textit{non}-resonant Raman spectrum. It is worth to mention that, in
the original work of Albrecht, a similar relationship has also been
discussed. However, it only focused on the case of near-resonance
conditions (see Eq.~22 in Ref.~\citenum{albrecht1961jcp}).

\section{Time-dependent Frame}

Time-dependent frame of Raman theory has the advantage of avoiding the
sum-over-states of $|v^r\rangle$ in Albrecht's
theory.\cite{lee1979jcp,heller1981acr,heller1982jpc,tannor1982jcp,lee1985jce}
Instead, the dynamics of the wave packet given by the initial
vibrational wave function times the corresponding electronic transition
dipole on the excited states is involved.\cite{lee1979jcp}
Mathematically, the equivalence of these two representations could be
readily proven by a simple half-Fourier transform.\cite{lee1979jcp,
heller1981acr} In time frame, the ``dynamic'' and ``static'' terms are
equivalent to the Albrecht's $A$ and $B$ terms, respectively. Here, the
``dynamic'' term arises from wave packet propagation on the
Born-Oppenheimer surfaces, while, the ``static'' term comes from the
coordinate dependence of the electronic transition
dipole.\cite{lee1979jcp} In other words, the first and second terms in
the Taylor expansion of transition dipole, \textit{i.e.}
$p_{e^re^g}=p_{e^re^g}^0+p_{e^re^g}^{k,0}Q_k$, associate with the
``dynamic'' and ``static'' terms, respectively.

In \textit{non}-resonant conditions, the short time approximation holds.
Thus, in the zero-frequency limit, the time average polarizability
associated with $|e^r\rangle$ could be calculated
as\cite{lee1979jcp,half_f}
\begin{equation}
\bar{\alpha}_{pq}^r=\frac{2\imath}{\hbar{T}}\int_0^T\alpha_{pq}^r(s){d}s,
\label{eq:apq_avt}
\end{equation}
where the per-factor $2$ arises from the zero-frequency limit,
$T=2\pi\hbar/|E_{\textrm{av}}-E_{e^gv^i}|$, and
\begin{equation}
\alpha_{pq}^r(s)=\int_0^s{d}t
\exp\left(\frac{-\imath{E}_{e^gv^i}t}{\hbar}\right)
\langle\phi_f|\exp\left(\imath\mathcal{H}_rt/\hbar\right)|\phi_i\rangle.
\label{eq:apq-s}
\end{equation}
Here $\mathcal{H}_r$ is the vibrational Hamiltonian for $|e^r\rangle$,
the damping factor is omitted, $\langle\phi_f|=\langle{v}^f|p_{e^ge^r}$,
$|\phi_i\rangle=q_{e^re^g}|v^i\rangle$, and $E_{\textrm{av}}$ is the
average energy on $|e^r\rangle$. If the
assumption\cite{lee1979jcp,tang1970}
\begin{equation}
\langle\phi_f|\mathcal{H}_r-E_{\textrm{av}}|\phi_i\rangle=0
\label{eq:hr-t}
\end{equation}
is adopted, the ``dynamic'' term in Eq.~\ref{eq:apq-s} would be
\begin{equation}
\alpha_{pq}^{r,D}(s)=\int_0^s{d}t
\exp\left(\frac{\imath\Delta{E}t}{\hbar}\right)p_{e^ge^r}^0q_{e^re^g}^0
\langle{v}^f|v^i\rangle,
\label{eq:apq-sd}
\end{equation}
where $\Delta{E}=E_{\textrm{av}}-E_{e^gv^i}$. As a result, the
``dynamic'' term (the $A$ term) of the Stokes peaks at fundamental
frequencies also vanishes in the time frame due to the orthogonality of
the vibrational wavefunctions in the ground state.  On the other hand,
the ``static'' terms survives and its value in the first order of $Q_k$
is exactly equal to $B$ term in Eq.~\ref{eq:ar} if we considered
$\Delta{E}=\Delta{E}_{rg}$ (see Appendix B in Ref.~\citenum{lee1979jcp}
for details). This conclusion is consistent with that in
Ref.~\citenum{lee1979jcp}, where the emphasis for far from resonance is
solely on the ``static'' term.

The reason of the vanishing of ``dynamic'' term in
\textit{non}-resonance in the time frame would be easily located as the
assumption of Eq.~\ref{eq:hr-t}, which is equivalent to the assumption
of $\omega_{e^rv^r:e^gv^i}\approx\omega_{e^r:e^g}$ in the energy frame.
Both assumptions neglect the vibrational contributions in excited
states. To obtain the correct result, in the time frame,
Eq.~\ref{eq:apq_avt} should be calculated with the inclusion of
vibrational contributions in $\mathcal{H}_r$, so that the contribution
of the ``dynamic'' term (Albrecht's $A$ term) will be included in the
total intensity.

\section{First principles example: Water monomer}

\begin{figure}[!tb]
\begin{center}
\includegraphics[width=8.6cm]{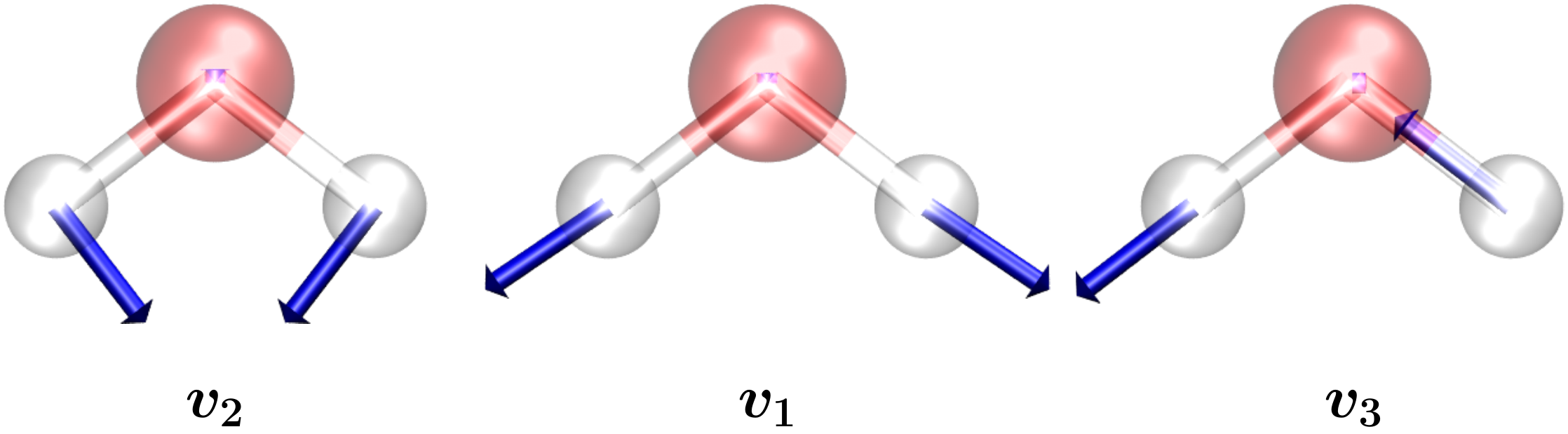}
\caption{\label{fig:vib}Three vibrational modes of water molecule.}
\end{center}
\end{figure}

Although the contribution of $A$ term for \textit{non}-resonant Raman
intensities has been briefly
touched,\cite{shorygin1947zfk,ting1968sa,lee1979jcp} its magnitude for
realistic molecules was still
controversial\cite{bishop1997prb,castiglioni1997prb}. Therefore, a first
principles example is highly desirable to address this issue. To the
best of our knowledge, there were no previous examples that have
investigated the magnitudes of both $A$ and $B$ terms at first
principles level for \textit{non}-resonant Raman intensities.  In the
following, we take water molecule as a realistic example to fully
examine the contribution of the $A$ term. Water molecule has three
vibrational modes,\cite{cotton1990} which are $\angle\textrm{H-O-H}$
bending ($v_2$), symmetric O-H stretching ($v_1$), and asymmetric O-H
stretching modes ($v_3$), respectively. The vibrational modes of water
are depicted in Fig.~\ref{fig:vib}. All required excited states except
core-hole excitations were calculated at time-dependent density
functional theory level.\cite{comp_method} The finite difference
method\cite{fornberg1988mc} was used to calculated the derivatives in
$\alpha_{pq}^r$. All calculated $\alpha_{pq}^r$ were summed up over all
singly excited states to obtain $\alpha_{pq}$ and finally Raman
scattering factors were calculated by\cite{jcc-raman,wilson1980}
\begin{equation}
S_k=45a_k^2+7\gamma_k^2,
\label{eq:S}
\end{equation}
where
\begin{equation}
\begin{split}
a_k=&\frac{1}{3}\left\{\left(\alpha_{xx}\right)_k+
\left(\alpha_{yy}\right)_k+\left(\alpha_{zz}\right)_k\right\}\\
\gamma_k^{2}=&\frac{1}{2}\bigg\{%
\left[\left(\alpha_{xx}\right)_k-\left(\alpha_{yy}\right)_k\right]^2\\
&+\left[\left(\alpha_{yy}\right)_k-\left(\alpha_{zz}\right)_k\right]^2\\
&+\left[\left(\alpha_{zz}\right)_k-\left(\alpha_{xx}\right)_k\right]^2\\
&+6\left[\left(\alpha_{xy}\right)_k^2+\left(\alpha_{xz}\right)_k^2
+\left(\alpha_{yz}\right)_k^2\right]\bigg\}.
\end{split}
\label{eq:Sab}
\end{equation}
Here the subscript ``$k$'' represents the individual vibrational mode.
By using Eq.~\ref{eq:ar1}, we have calculated the scattering factors
from three cases: only the $A$ term ($S_A$), only the $B$ term ($S_B$),
and both $A$ and $B$ terms ($S_{\textrm{Tot}}$). With these scattering
factors, the corresponding Raman intensities, \textit{i.e.}, $I_A$,
$I_B$ and $I_\textrm{Tot}$, are calculated
as\cite{long2002,jcc-raman}
\begin{equation}
\begin{split}
I=\sum_k&\frac{\pi^2}{\epsilon_0^2}
\left(\tilde{\nu}_{\textrm{in}}-\tilde{\nu}_k\right)^4\\
&\times\frac{h}{8\pi^2c\tilde{\nu}_k}
\frac{S_k}{45}\frac{1}{1-\exp\left(-hc\tilde{\nu}_k/k_BT\right)},
\label{eq:dcs}
\end{split}
\end{equation}
where $\epsilon_0$ is the vacuum permittivity,
$\tilde{\nu}_{\textrm{in}}$ and $\tilde{\nu}_k$ are the wave number of
incident light and vibrational mode, respectively, $h$ is the Planck's
constant, $c$ is the speed of light, $k_B$ is the Boltzmann constant,
and $T$ is the temperature. By definition, the interference contribution
is calculated by $I_{\textrm{Tot}}-I_A-I_B$. For comparison, the
analytical Raman scattering factors at the same density functional
theory level were also computed.

\begin{table}[!tb]
\caption{Calculated Raman scattering factors for water (in the unit of
$\textrm{\AA{}}^4/\textrm{amu}$) contain only the $A$ term ($A$), the
$B$ term ($B$), or all terms ($\textrm{Tot.}$), as well as analytical
results (Anal.). The experimental measurements (Expt.) extracted from
Refs.~\citenum{amos1986cpl}, \citenum{murphy1977mp}, and
\citenum{murphy1978mp} also included for comparison.}
\label{tab:raman}
\begin{ruledtabular}
\begin{tabular}{ccrrrrr}
 Mode  & Symmetry & \multicolumn{1}{c}{$A$}
 & \multicolumn{1}{c}{$B$}   & \multicolumn{1}{c}{$\textrm{Tot.}$}
 & \multicolumn{1}{c}{Anal.} & \multicolumn{1}{c}{Expt.} \\
\hline
  $v_2$ & $A_1$ &  0.14  &  1.54   &   1.10  &   1.10  &  0.9$\pm$0.2 \\
  $v_1$ & $A_1$ & 11.66  & 48.32   & 104.94  & 104.54  &  108$\pm$14  \\
  $v_3$ & $B_2$ &  0.00  & 25.68   &  25.68  &  25.67  & 19.2$\pm$2.1 \\
\end{tabular}
\end{ruledtabular}
\end{table}

\begin{figure}[!tb]
\begin{center}
\includegraphics[width=8.6cm]{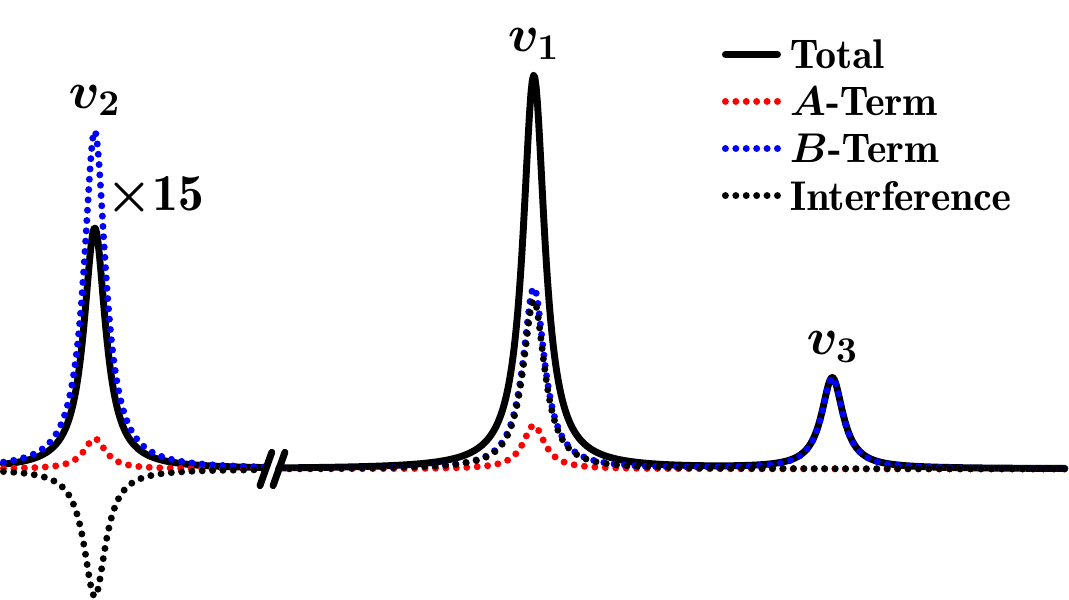}
\caption{\label{fig:int}Calculated Raman intensities for water (black
solid line). The contributions from the $A$ term (red dotted), the $B$
term (blue dotted), and the interference (black dotted) were also shown.
The incident light and temperature were set to be 514.5~nm and 400~K
followed by the experimental conditions,\cite{murphy1977mp,murphy1978mp}
respectively. All spectra have been broadened by Lorentzian function
with a full with at the half-maximum of 10~cm$^{-1}$.}
\end{center}
\end{figure}

All calculated scattering factors as well as experimental
values\cite{murphy1977mp,murphy1978mp} are listed in
Table~\ref{tab:raman}.  As expected, the calculated $S_{\textrm{Tot}}$
are identical to the analytical results and in good agreement with the
experimental values\cite{murphy1977mp,murphy1978mp,amos1986cpl}. Here a
slight overestimation for $v_3$ should be attributed to the approximate
exchange-correlation functionals, since, at the CASSCF(10,10) level with
the same basis set used here, the calculated value
(20.4~$\textrm{\AA{}}^4/\textrm{amu}$) is in the range of experimental
observation.\cite{neugebauer2002jcp} For the $v_3$ mode that belongs to
$B_2$ symmetry, the $A$ term does not contribute to the total intensity,
consistent with the discussion mentioned above.  For the $v_1$ and $v_2$
modes, on the other hand, the contribution of the $A$ term can no longer
be neglected. For instance, the calculated $S_B$ of $v_1$ is less than
half of the analytical value. It is interesting to note that $S_A$ can
only account for about 10\% of the analytical result and the summation
of contributions from $S_A$ and $S_B$ terms can not provide the correct
answer neither. On the other hand, a simple estimation shows that the
value from $(\sqrt{S_A}+\sqrt{S_B})^2$
(107~$\textrm{\AA{}}^4/\textrm{amu}$) can reproduce the analytic result
nicely. It indicates that a strong constructive interference between the
$A$ and $B$ terms does exit for the $v_1$ mode and covers almost 40\% of
the total intensity. A minor destructive interference is identified for
the $v_2$ mode. In this case, the simple summation of $S_A$ and $S_B$
gives a value larger than the final one. Contributions from different
terms for the Raman intensities are also displayed in
Fig.~\ref{fig:int}, which emphasizes the importance of the interference
for the relative intensity. Notice that the first excitation energy of
water is around 7~eV, which obviously obeys the condition of
$\omega_{e^rv^r:e^gv^i}\approx\omega_{e^r:e^g}$. But even though, the
contribution of the $A$ term cannot be neglected. Overall, although the
water monomer is a simple case, it is fully adequate to address our
argument.

We notice that simple extension of Eq.~\ref{eq:ar1} by replacing
$\Delta{E}_{rg}^2$ and $\Delta{E}_{rg}$ with
$2[\Delta{E}_{rg}\pm\hbar(\omega+\imath\Gamma)]^2$ and
$2[\Delta{E}_{rg}\pm\hbar(\omega+\imath\Gamma)]$,
respectively,\cite{gamma_long} was employed to calculate
dynamic\cite{carole2000pccp} and resonant Raman\cite{rappoport2011jpcl}
intensities.  Here the damping factor $\Gamma$ represents the life time
of the excited state. In addition, the same extension of the second term
(the $A$ term, labeled as $A_{pq}^{r,s}$ hereafter) in Eq.~\ref{eq:ar1}
has been used to explain the chemical enhancement of surface-enhanced
Raman spectroscopy.\cite{jensen2008csr,zayak2011prl} However, the
accuracy of such extension is questionable because, near the resonant
condition, the Taylor expansion for denominator in Eq.~\ref{eq:ab} may
fail and thus $|v^r\rangle\langle{v}^r|$ cannot be summed
prior.\cite{prio_sos}

\begin{figure}[!tb]
\begin{center}
\includegraphics[width=8.6cm]{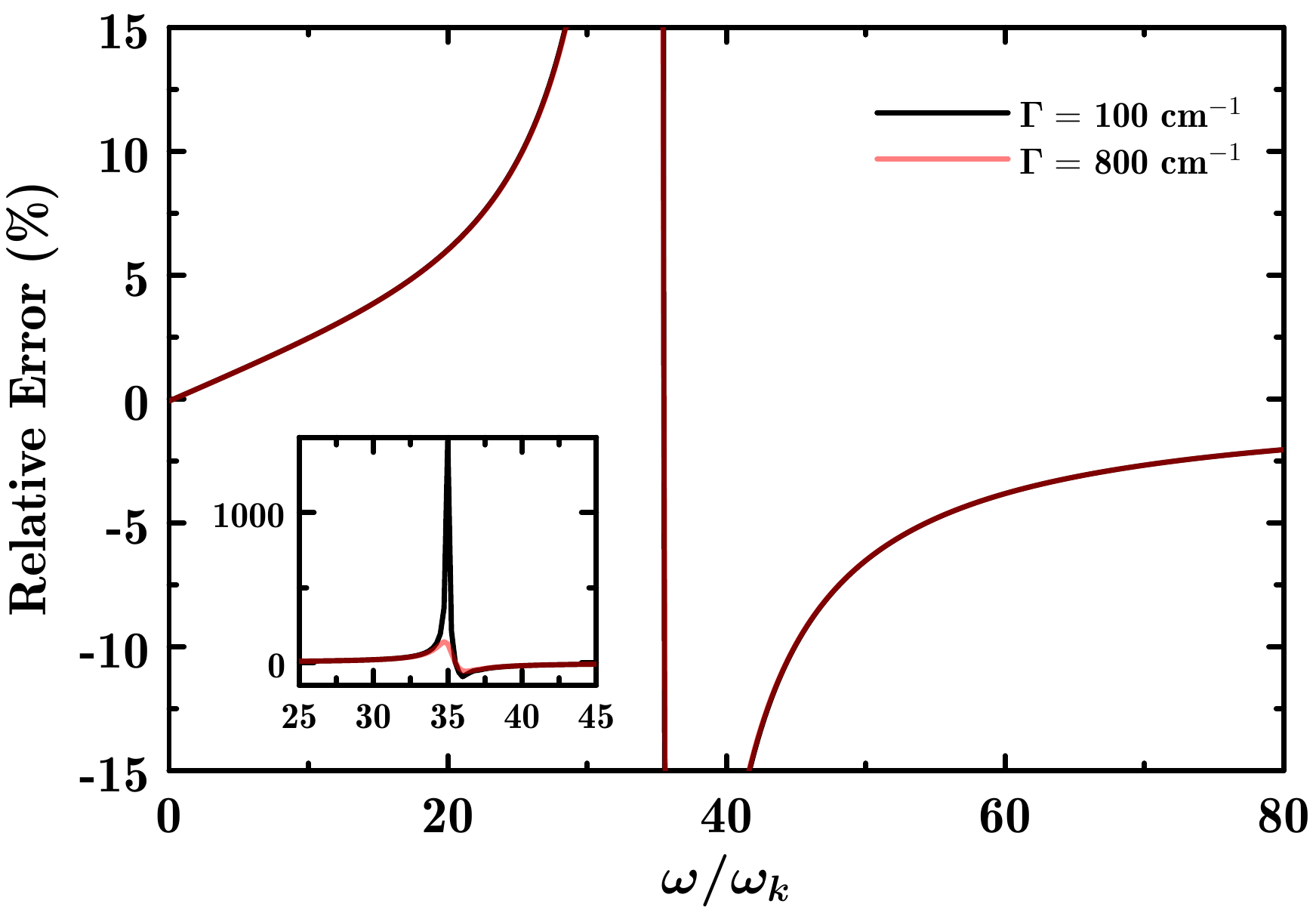}
\caption{\label{fig:comp}Relative error of simply extended $A$ term in
Eq.~\ref{eq:ar1} respect to Eq.~\ref{eq:ab} at different energy of
incident light for the model in Fig.~\ref{fig:model}.  Here all
parameters are rationalized from the first excited state of water
molecule with the bending mode in LCM.  Specifically,
$\Delta{Q}=1.35~\textrm{a.u.}$,
$\omega_k=\omega_k^\prime=1600~\textrm{cm}^{-1}$, and
$\omega_{e^r:e^g}=35~\omega_k$. Two $\Gamma$ of 100 and 800~cm$^{-1}$
were used. The insert figure shows the incident light around the
resonant situation.}
\end{center}
\end{figure}

To verify the accuracy of this simplified extension of Eq.~\ref{eq:ar1},
we computed the relative error of $A_{pq}^{r,s}$ respect to
Eq.~\ref{eq:ab} at different energy of incident light for the model
system shown in Fig.~\ref{fig:model}. The result is depicted in
Fig.~\ref{fig:comp}. Here all parameters were rationalized from the
first excited state of water molecule with the bending mode in LCM (see
caption of Fig.~\ref{fig:comp} for details).  Two typical $\Gamma$,
\textit{i.e.} small one of 100~cm$^{-1}$ and large one of
800~cm$^{-1}$,\cite{liang2012jctc,zhao2006jacs} were used. Around the
resonant situation, different $\Gamma$ introduces different behaviour
for the relative error. For instance, the maximum relative error for
small $\Gamma$ is much larger than that for large $\Gamma$.  For the
specific case, the maximum relative errors are even larger than 1400\%
and 130\% for small and large $\Gamma$, respectively.
In the region of away from resonant condition, the identical behaviour
for relative errors with different $\Gamma$ is observed in
Fig.~\ref{fig:comp}. Due to the uncertainty of experimental measurements
as well as the limitation of the double harmonic
approximation,\cite{dft-koch} we used 15\% as the threshold for the
calculated $A_{pq}^{r,s}$. According to the model calculations, we have
found that $A_{pq}^{r,s}$ is a good approximation when
$|\omega_{e^r:e^g}-\omega|$ is larger than $5~\omega_k$. We should
emphasize that this value is mainly determined by $\Delta{Q}$ between
two PESs. For example, when $\Delta{Q}=10~\textrm{a.u.}$, the range of
accurate $A_{pq}^{r,s}$ is $|\omega_{e^r:e^g}-\omega|>10~\omega_k$.
Apparently, when $\Delta{Q}=0$, the parity symmetry will lead to zero
$A$ term both in Eq.~\ref{eq:ab} and Eq.~\ref{eq:ar1}.

\section{Conclusion}

We have clearly shown that the contributions from the Albrecht's $A$
term are too large to be negligible for the \textit{non}-resonant Raman
intensities of totally symmetric fundamental vibrational modes, in
contrast with the common wisdom of the field. The widely employed
perturbation method is also found to be too crude for evaluating dynamic
or resonant Raman spectra.  Our findings are conceptually important for
correctly understanding and modeling of Raman scattering processes under
different conditions.    

\section*{Acknowledgments}

This work was supported by the Ministry of Science and Technology of
China (2010CB923300), the ``Strategic Priority Research Program'' of the
Chinese Academy of Sciences (Grant XDB01020200), the Natural Science
Foundation of China (21421063), and Swedish Research Council (VR). The
Swedish National Infrastructure for Computing (SNIC) was acknowledged
for computer time.


\begin{thebibliography}{50}
\bibitem [{com()}]{comp_method}%
  \BibitemOpen
  \href@noop {} {}\bibinfo {note} {{The used functional and basis set were
  B3LYP implemented in Gaussian 09 and Sadlej pVTZ, respectively.}}\BibitemShut
  {Stop}%
\end{thebibliography}
%

\end{document}